# 3D Holographic Flow Cytometry Measurements of Microalgae: Strategies for Angle Recovery in Complex Rotation Patterns

**Francesca Borrelli,**[a] **Giusy Giugliano,**[a,b] **Emilie Houliez,**[c] **Jaromir Béhal,**[a] **Daniele Pirone,**[a] **Leonilde Roselli,**[c,d] **Angela Sardo,**[c] **Valerio Zupo,**[e] **Maria Costantini,**[c] **Lisa Miccio,**[a] **Pasquale Memmolo,**[a] **Vittorio Bianco,**[a,*] **Pietro Ferraro**[a]

[a]Institute of Applied Sciences and Intelligent Systems "E. Caianiello", Consiglio Nazionale delle Ricerche (ISASI-CNR), via Campi Flegrei 34, Pozzuoli (NA), Italy.

[b]Department of Mathematics and Physics, University of Campania, Caserta 81100, Italy.

[c]Stazione Zoologica Anton Dohrn, Villa Comunale 80121 Napoli, Italy

[d]NBFC, National Biodiversity Future Center, Piazza Marina 61, 90133 Palermo, Italy

[e]Dipartimento di Biotecnologie Marine Ecosostenibili, Stazione Zoologica Anton Dohrn, Ischia (NA), Italy.

[*]vittorio.bianco@isasi.cnr.it

**Abstract**

Marine ecosystems are in the spotlight, because environmental changes are threatening biodiversity and ecological functions. In this context, microalgae play key ecological roles both in planktonic and benthic ecosystems. Consequently, they are considered indispensable targets for global monitoring programs. However, due to a high spatial and temporal variability and to difficulties of species identification (still relying on microscopy observations), the assessment of roles played by these components of marine ecosystems is demanding. In addition, technologies for a 3D assessment of their complex morphology are scarcely available. Here, we present a comprehensive workflow for retrieving 3D information on microalgae with diverse geometries through holographic microscopy operating in flow-cytometry mode. Depending on the rotation patterns of samples, a tailored approach is used to retrieve their rolling angles. We demonstrate the feasibility of measuring 3D data of various microalgae, contingent to the intrinsic optical properties of cells. Specifically, we show that for quasi-transparent and low-scattering microorganisms, the retrieved angles permit to achieve quantitative 3D tomographic Refractive Index (RI) mapping, providing a full characterization of the alga in terms of its inner structure and the outer shape. Moreover, even in the most challenging scenarios, where microalgae exhibit high light absorption or strong scattering, quantitative 3D shape reconstructions of diatoms and dinoflagellates can be at least achieved. Finally, we compare our direct 3D measurements with 2D inferences of 3D properties, obtained using a commercially available microscopy system. The ability to non-invasively obtain 3D information on microalgae marks a fundamental advancement in the field, unlocking a wealth of novel biological insights for characterizing aquatic ecosystems.

# 1 Introduction

Concerns about environmental changes trigger major challenges, in the modern era, because human activities significantly impact all ecosystems on Earth, making their preservation a top priority. In this context, particular attention is given to the functioning of marine ecosystems, that depends on their structure and biodiversity [1]. Among the biological components that are considered of primary importance, microalgae play a crucial ecological role, contributing to nearly half of the planet's primary productivity, forming the foundation of food webs, and affecting the biogeochemical cycles of key organic compounds [2]. Human-induced processes greatly affect the health and the community organization of primary producers, making them valuable bioindicators for monitoring the status of aquatic ecosystems [3]. Due to their prominent ecological role, microalgae are considered an indispensable target for global monitoring programs [4] and have traditionally been worldwide included in routine assessments by environmental agencies, as descriptors of the ecological status of aquatic ecosystems [5]. However, the high spatial and temporal variability of microalgae and the difficulties intrinsic into species identification activities, synergistically contribute to rend a reiterated assessments of these ecological components particularly demanding [6]. Species identification is based mainly on inverted light microscopy [7], electron microscopy and DNA sequence analysis [8, 9]. These techniques are typically time-consuming, costly, and limiting in efficiency [10]. In addition, the classical manual microscopy is often necessarily integrated with scanning or transmission electron microscopy in order to obtain the diagnostic features of cell structure and ultrastructure. The increasing need for rapid, cost-effective tools to characterize microalgae communities in a time-efficient way (collection, identification and quantification) is gradually going towards high throughput imaging applied *in situ* or on collected samples [11]. Although the application of machine learning algorithms gives encouraging results in classifying the heterogenous microalgal community, there is still a lack of technology for the three-dimensional (3D) investigation of the complex morphology of these organisms. As a consequence, some morphological descriptors of microalgae, such as biovolume - a critical parameter for biomass calculation - remain rough estimations. This is a serious gap considering the crucial roles morphology plays in the survival and ecological interactions of microalgae. For instance, morphology influences predator avoidance, nutrients uptake, light absorption, buoyancy and motility [12]. Furthermore, anthropogenic activities can have a deep impact on the morphology of microalgae, potentially making morphology a good bioindicator for pollution.

Digital Holography (DH) has emerged as a promising alternative to conventional microscopy. Among the advantages offered, Digital Holography (DH) is a quantitative, label-free technique with the ability to numerically refocus images after acquisition and to capture the entire complex field modulated by the sample. This makes it very flexible and powerful, as it can image samples in correct focus post-holographic acquisition [13]. The reconstruction process of digital holograms furnishes at the same time amplitude as well as the so-called Quantitative Phase Maps (QMPs) that represent the quantitative integral measurements of the optical path delay introduced by the sample, when a coherent laser wavefront is transmitted through it. One possibility offered by DH is also the ability to reconstruct the 3D RI profile via tomography. Among others, In-Flow Holographic Tomography (FHT) is a breakthrough that exploits the sample rotation induced in microfluidic channels, so that the illumination is kept fixed while the object rolls and gets probed from multiple directions [14]. However, one challenging issue is the accurate retrieval of rotation angles to be associated to each QPM. Recently, it has been demonstrated that the retrieval of rotation angles can be obtained for spherical or quasi spherical samples (such as, suspended cells) under the assumption of uniform rotation along a single axis [15] (see Supplementary Information). Applying this approach to microalgae is challenging because of the difficulties associated with their non-spherical morphology, interclass heterogeneity and much more complex rotation patterns experienced by non-spherical objects. Furthermore, among the extremely vast diversity of algae, some exhibit low light scattering and absorbance, while others exhibit non-negligible scattering and light absorbance.

Thus, the former can be characterized by FHT, while the latter cannot, as the QPMs cannot be extracted. Nonetheless, even in the cases of algae having strong scattering and/or non-negligible light absorbance, the holographic technology combined with flow cytometry permits retrieving 3D information that is meaningful for ecosystem studies. In fact, in the latter cases, valuable information for classification, morphometric evaluations and biovolume estimation can be recovered by using algorithms such as the Shape From Silhouette (SFS) [16, 17]. Essentially, DH permits to reconstruct amplitude and phase of the diffracted wavefront and we used refocused Amplitude Maps (AMs) for the SFS recovering. Regardless of the technique exploited to recover the 3D shape (FHT or SFS), the in-flow 3D cytometry requires an automated and reliable approach to recover the rolling angles sequence for each microalga.

Here, we show a novel workflow designed for tracking the angular sequences of non-spherical objects in continuous flow. The block-scheme is sketched in Figure 1. We show that a different approach needs to be developed to reconstruct its three-dimensional structure, depending on the specific shape of the microalga, of its optical properties and the type of rotation it performs. We demonstrate, as well, that a single holographic optical instrument can provide quantitative 3D measurements of the tested sample, regardless of the microalgae's intrinsic optical properties. Moreover, we discuss the discrepancy between the proposed approach and estimations obtained with the Imaging Flow CytoBot (IFCB). Such commercial imaging flow cytometers are typically used to image and characterize microalgae in 2D and thus have inherent limitations in retrieving 3D morphological features from raw 2D measurements. It is important to note that this work introduces a novel strategy for measuring angle rotations that allows for retrieving FHT as well as SFS. However, a full analysis of the 3D distribution of a species from tomographic data is beyond the scope of this study. Here, we restrict our analysis to the 3D outer shape of the microalgae, focusing on morphometric evaluations and biovolume estimation. Such parameters are common data for techniques like FHT and SFS. Future studies will explore the analysis of inner structures from the 3D RI tomograms of transparent and weak scattering microalgae, made viable after having established a robust angle retrieval workflow.

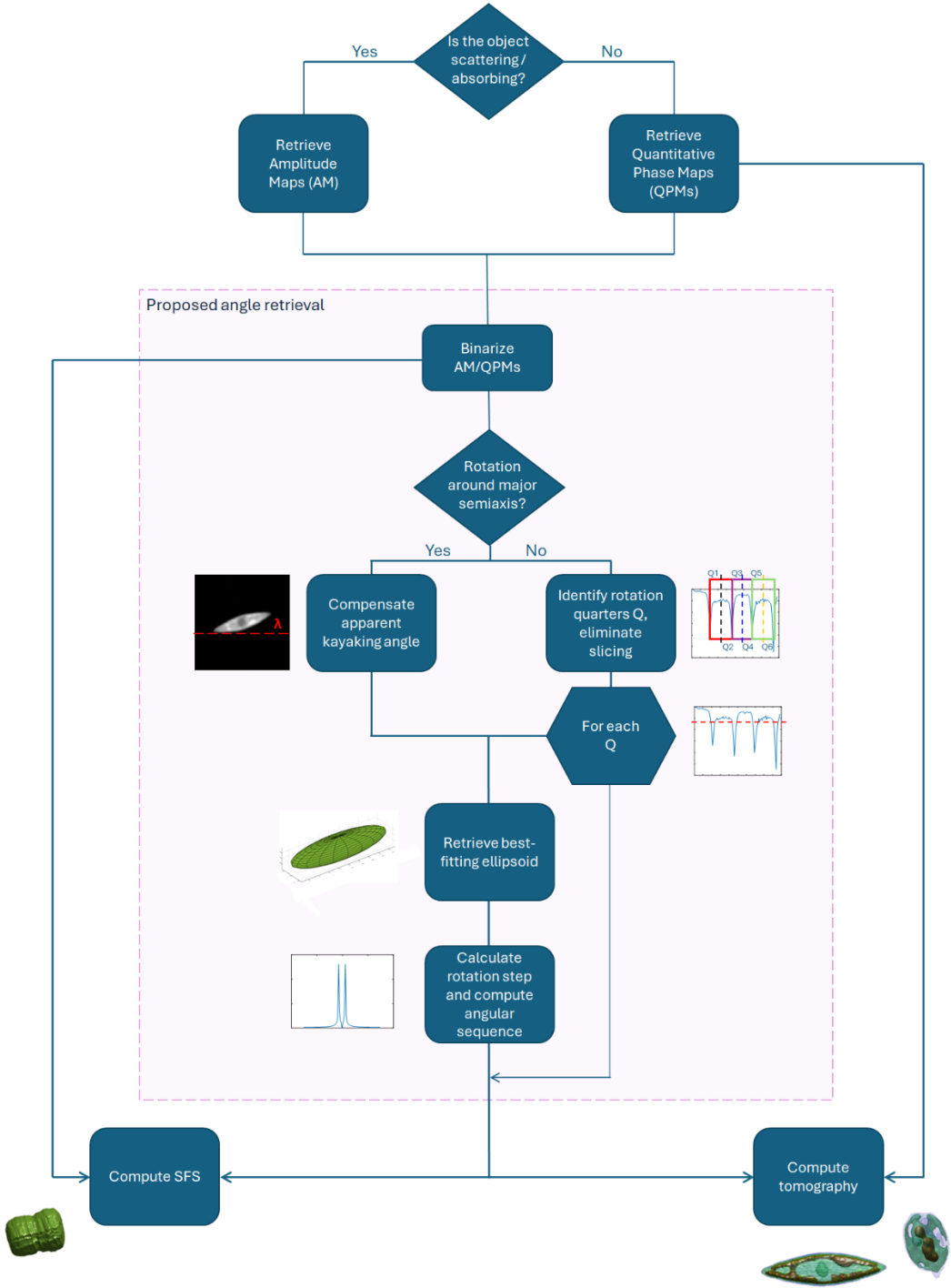

**Figure 1:** Workflow for the proposed angle retrieval pipeline for non-spherical samples. Depending on the specific rotation pattern followed by the sample during its rotation and by its optical properties, different strategies are used for the recovery of its 3D shape and internal structures.

## 2 Materials and Methods

*2.1 Rotation of non-spherical samples*

To develop a correct angle retrieval method for non-spherical samples, we previously analysed their rotation pattern in the microfluidic setup in use. We observed the rotation of all samples in a laminar flow, which is complex and varies among algal species in relation to their shape and entering orientation. Accurate mathematical descriptions have been proposed in [18-20] for a spheroid, which is a rotational solid. According to the pivotal study [18], the rotation has components along all the three cartesian axes x, y, z, which complicates the 3D recovery. Therefore, it is necessary to quantify and possibly compensate the spurious rotations and then proceed with reconstruction. The possibility to compensate the spurious rotations depends on whether the sample is rotating around its major or minor semiaxis, *i.e*, which profile it offers with respect to the acquisition plane. Since this differentiation is of major importance in the proposed angle retrieval method, throughout this paper we will refer to the rotation happening around the major or minor semiaxes as "Rotation pattern 1" and "Rotation pattern 2", respectively, as sketched in Figure 2. Another consideration is due for approximate ellipsoidal samples, characterized by three different semiaxis values. In this case, the angular velocity changes whether the sample faces the flow with its major or minor cross-section. A frequent condition is that the sample exhibits a sharp rotation when its major cross-section is perpendicular to the flow. In contrast, when its major cross-section is aligned to the flow, the sample experiences a sliding along the microfluidic channel before completing its rotation, because the torque is insufficient to induce a rotation (tumbling-sliding alternance; Figure 2).

In fact, we observed that the sliding of the sample stops, resulting in a quick 180° flip, when a sufficient cross-section is exposed to the flow. This occurs due to the induced torque, which depends on the sample's minor cross-section and the velocity profile within the microfluidic channel. Furthermore, we frequently observed that the samples did not undergo a complete 360° rotation in the holographic microscope's imaging region, as no perturbation was present to modify the sliding phase. This phenomenon tends to occur more frequently for samples following Rotation pattern 2, as the two cross sections have hugely different areas. Lastly, the intrinsic optical properties of microalgae warrant consideration. Indeed, as already said above, certain species characterized by elevated absorption or scattering cannot be treated as a pure phase, weakly scattering samples, and thus are more appropriately described by their amplitude images rather than their quantitative phase-contrast maps. However, DH gives access to the entire complex field of the object. For each microalgal species, the most reliable component can be selected and extracted (i.e., QPMs or AMs) (see Supplementary material for further details). In the approach proposed here, the angles estimations are based on geometrical considerations and on the use of the binarized projection of the QPMs / AMs. In the following we name Binary Maps (BMs), such areas for proceeding with the most appropriate method for 3D reconstruction (i.e., RI tomography or SFS).

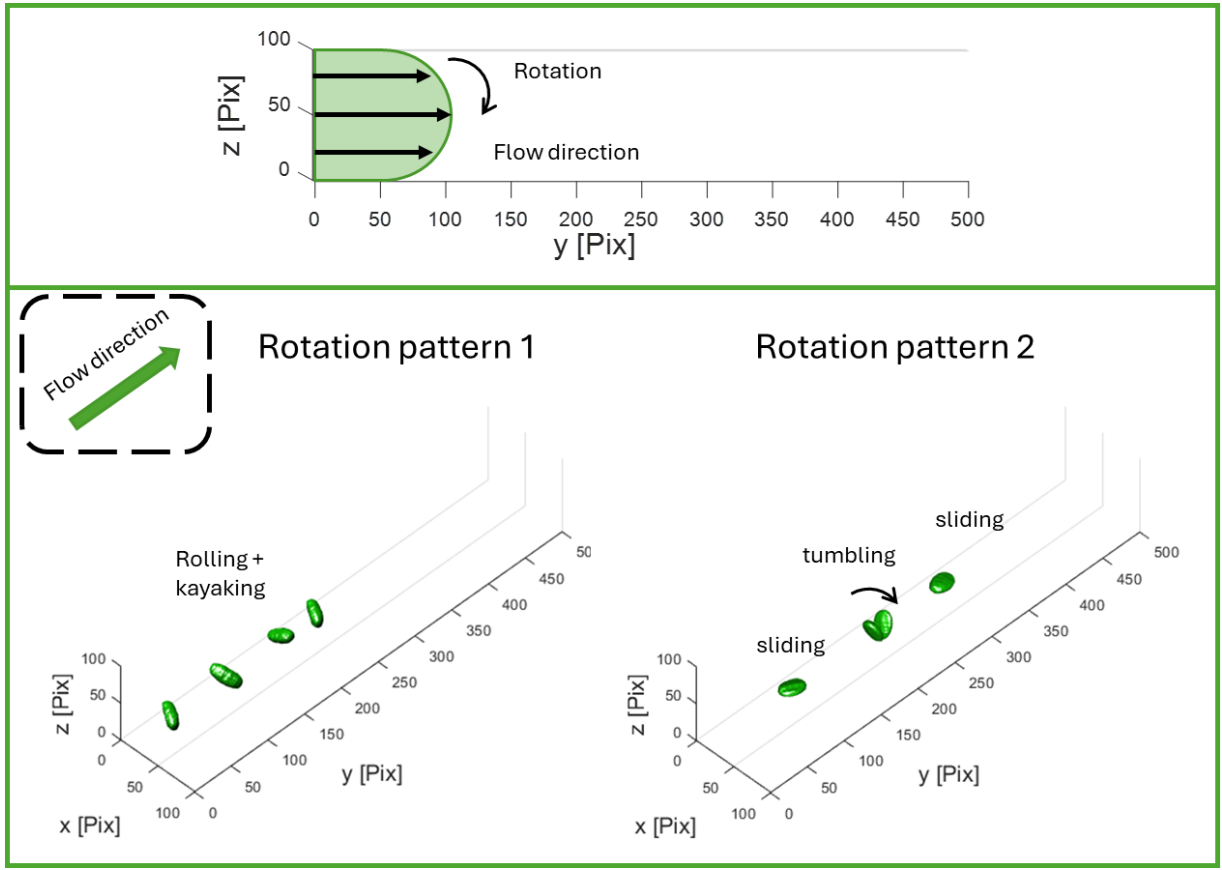

**Figure 2:** Illustration of the geometry of our microfluidic system and schematization of the nomenclature used for the rotation patterns discussed in this work. Upper panel, velocity profile in the microfluidic channel. Lower panel, schematization of the terminology introduced in this work. In Rotation pattern 1 the sample rotates around their major semiaxis, while in Rotation pattern 2 the sample rotates around a minor direction, resulting in the peculiar sliding-tumbling alternance.

*2.2 Angle retrieval in Rotation pattern 1*

As mentioned in the previous section, the preliminary step to recover the angular projections is to identify the spurious rotation contributions. Referring to the reference system presented in Figure 3a, the rotation component of our interest for the 3D reconstruction is the one happening around the x axis. The additional components around y and z are spurious. We define the BM planes as the family of planes perpendicular to the z axis. The motion around the y axis of an angle α happens in the xz plane, perpendicular to the BMs planes, as sketched in Figure 3b. This movement reflects in the BM planes as a variation of the length of the projection of the major semiaxis, $a$, resulting in a change in the focusing distance. Figure 3c plots the value of $a$ for the diatom *Navicula* sp. The maximum value of $a$ ($a_{max}$) corresponds to the condition for which the oscillation angle, α, is equal to 0, i.e, the sample lies in the BM planes and $a$ corresponds to the real major semiaxis length of the sample. The minimum values of $a$ ($a_{min}$) correspond to the maximum of α and represent the points in which the sample has the biggest tilt with respect to the BM planes. The value of α can be estimated recalling that $a_{min} = a_{max} \cos \alpha$. Figure 3d represents the evolution of α for the diatom *Navicula* sp., and the maximum value of α was 18.7°. The maximum relative variation of $a$ measured as $\frac{(a_{max} - a_{min})}{a_{max}}$ is equal to 0.053 μm. Since this oscillation occurs in the xy plane, it is not possible to compensate it, thus leading to a certain approximation in the 3D reconstruction (see Supplementary material). The second spurious rotation component occurs around the z axis. This component determines an oscillation of the BMs with respect to the horizontal direction, a phenomenon that we herein name as *apparent kayaking*, sketched in Figure 3e. It is possible to estimate the *apparent kayaking* angle λ evaluating the angle between $a$ and the x direction, and compensate it by rotating the BM of an angle λ towards x. In Figure 3g we plot the evolution of λ over time for *Navicula* sp., while in Figure 3f we show the effects of the *apparent kayaking* over the retrieved BM. Once the two spurious rotation contributions have been accounted for, the angular projections can be estimated starting from the *apparent kayaking* compensated BMs. To do so, we propose an approach inspired by [21, 22], presented in Figure 4 for a sample of *Navicula* sp.

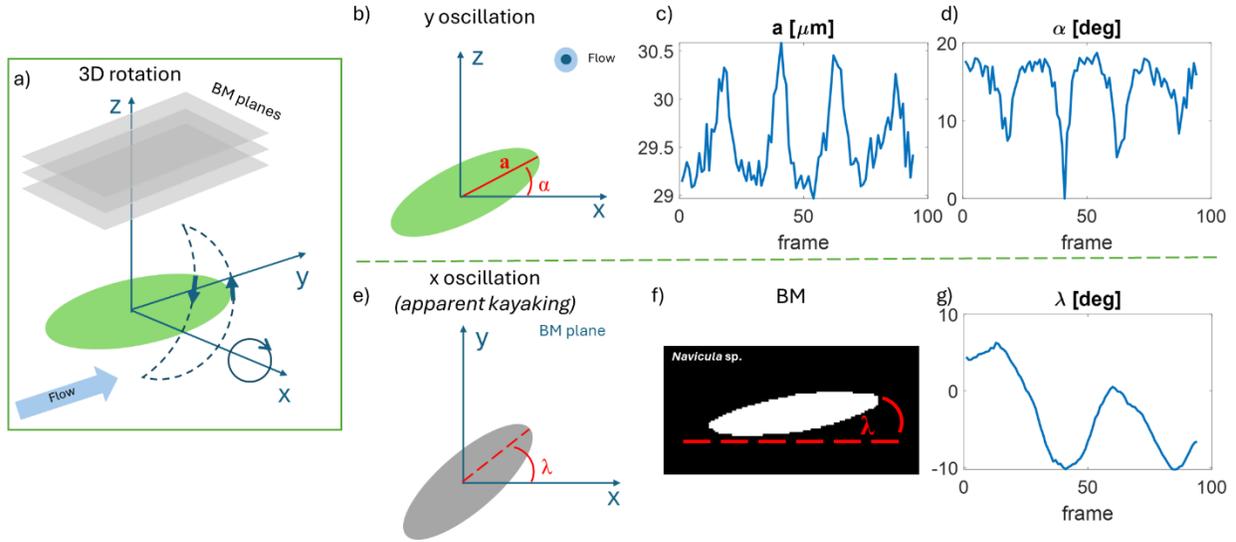

**Figure 3:** Illustration of the spurious rotation components in Rotation Pattern 1. (a), visualization of the rotation movement in the 3D space. (b), sketch of the y-spurious component rotation in the 3D space, defining $\alpha$ as the angle between the major semiaxis and the x asis. (c), measurement of the length of the major axis for a sample of *Navicula* sp., where the change due to the y-spurious rotation component can be appreciated. (d) evaluation of the angle $\alpha$ for a sample of *Navicula* sp. (e), definition of the apparent kayaking angle $\lambda$ in the BM planes. (f), identification of $\lambda$ over a real BM. (g) evaluation of $\lambda$ for a sample of *Navicula* sp.

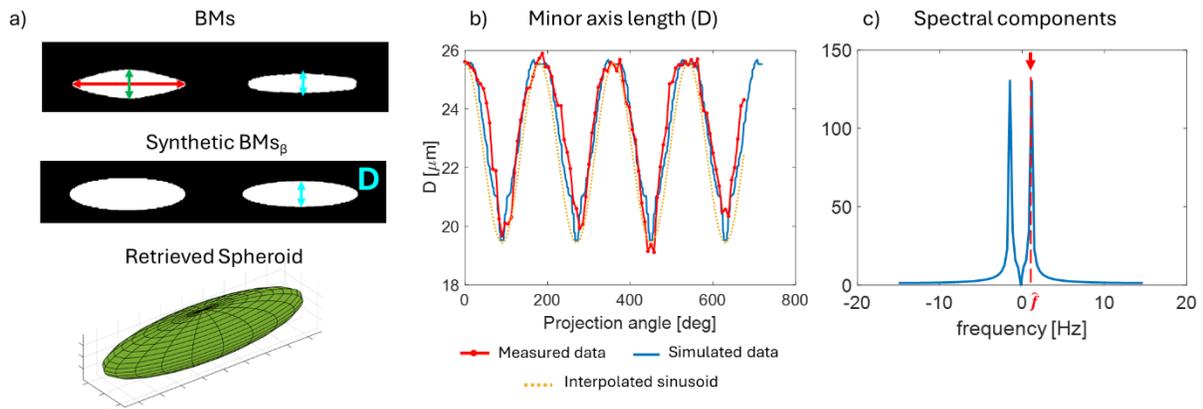

**Figure 4:** proposed angle retrieval method for non-spherical samples for Rotation pattern 1. (a), from the BMs the dimensions of the sample are measured and used to retrieve the spheroid representing it. From the numerical rotation and integration of the spheroid the synthetic BMs$_\beta$ are synthetized. (b), Value of the minor axis length D along the real BMs (red line), the BMs$_\beta$ (blue line) and the interpolated sinusoid (yellow line). (c), spectral component of the interpolated sinusoid, from which the angular step is computed.

Figure 4a, top, shows two BMs of a *Navicula* sp. during its rotation after the correction of the *apparent kayaking*. Starting from the BMs, the dimension of the sample in the three cartesian directions is measured. A spheroid of the corresponding dimension is then synthesized, which models the specimen, sketched in Figure 4a (bottom); following, a complete and exhaustive set of numerical phase maps $BMs_\beta$ is generated by projecting the spheroid along the angular sequence $\bar{\beta} = [\beta_1 ... \beta_N]$, obtained choosing an angular step of 0.01 degrees and sketched in Figure 4 a, centre. The synthetic step is strongly inferior to the expected real one (linked to the framerate of acquisition). From all the $BMs_\beta$ the value of the minor semiaxis D is measured, from which the angular frequency of the rotation will be calculated. Plotting D for all the simulated projections yields a harmonic curve, depicted in blue in Figure 4b. The corresponding curve measured from the real data is superimposed in red as a comparison. Starting from the curve representing the variation of D, its sinusoidal-fitted corresponding curve $y(t) = A \sin 2\pi \hat{f} t$ is obtained, depicted also in Figure 4b with a dashed yellow line. In this equation, $A$ is the amplitude of the fitted sinusoid, while $\hat{f}$ is its oscillation frequency. It must be noted that the oscillation frequency $\hat{f}$ is double compared to the rotation frequency of the spheroid, as D reaches its maxima twice during a full rotation of the spheroid, respectively for 180° (half rotation) and 360° (full rotation). In formulas, naming $f, T$ the frequency and the period of the rotation of the spheroid respectively, we obtain $f = \frac{\hat{f}}{2}, T = 2\hat{T}, \hat{T} = \frac{1}{\hat{f}}$. By observing the amplitude Fourier spectrum of this fitted curve after its mean value, $m$, is subtracted, $|Y(f)| = |\mathcal{F}\{y(t) - m\}|$, two peaks appear, located respectively at $\pm \hat{f}$, as reported in Figure 4c. Accordingly, the angular pulsation for the spheroid rotation can be expressed as $\omega = 2\pi f = 2\pi \frac{\hat{f}}{2} = \pi \hat{f}$ rad/s, which corresponds to $180\hat{f}$ deg/s. Finally, the sought-after angular step is obtained as $angular\ step = \frac{180\ \hat{f}}{F_s}$, where $F_s$ is the camera framerate. Once the angular step is known, the entire angular sequence can be retrieved by assigning to the first BM the angle 0°. The sinusoidal interpolation of the variation of D along the BMs corresponds to a frequential filtering of the angular speed. Indeed, due to the slight asymmetry of real samples which are not ideal spheroids, the angular speed might not be homogeneous during the period, resulting in an altered sinusoid and therefore spurious frequential component. These contributions are filtered by our method. However, as experimentally demonstrated, since the samples discussed in this case exhibit small asymmetries, this approximation does not alter the final reconstruction.

*2.3 Angle retrieval for Rotation pattern 2*

The rotation pattern 2 represents the most challenging condition in our experimental setup due to the presence of the tumbling-sliding sequence. As we did for the previous class, the first step is to correct the spurious rotation contributions around the y and z axes. The rotation around the z axis, which determines the *apparent kayaking*, can be evaluated in the sliding phase, still defining it as the angle between the horizontal direction and the major semiaxis. However, due to the tendency of the samples to align with the flow direction in these rotation cases, this angle tends to be larger than 45°. Therefore, the correct compensation should be made by rotating the sample towards the y direction, to keep the rotation happening around the minor semiaxis, of a quantity equal to $90 - \lambda$. This definition applies only to the BMs in the sliding phase, as when the sample is tumbling the major semiaxis of the recorded BMs does not coincide with the specimen's one. Indeed, the definition stated above for the *apparent kayaking angle* does not correspond to $\lambda$ for these BMs. The correct compensation should be made in the xz plane, which is perpendicular to the BM planes and therefore not possible in our experimental setup. However, since the thumbling phase is generally fast, this correction can be neglected.

For the angular sequence estimation, a serious concern is also related to the automatic identification of a periodicity in the rotation. The approach we propose for the automatic identification of the rotation frames is summarized in Figure 5. Figure 5a represents the application of the method in [15] to a typical spherical sample. The clear alternance between maxima and minima in the correlation coefficient helps to assign the 360° frame and to identify the BMs assigned to 90°, 180° and 270° for all the periods recorded. The correlation coefficient leverages the phase values to find correspondences, as no consideration can be made on the geometry. The angle retrieval pipeline is completely automatic, provided that the full rotation BM can be identified. However, the application of this procedure to a non-spherical sample (in this example, we used *Cocconeis* sp. in Rotation pattern 2.) would result in an ambiguity, as reported in Figure 5b. According to this plot, multiple BMs maximise the similarity metrics with respect to the first orientation, all relative to the sliding phase, thus making impossible to apply the automated procedure typically adopted for spherical samples. According to the method we propose, the starting point is the evaluation of the area of the BMs, as plotted in Figure 5c. From this graph the points of minimum area are clearly recognizable, corresponding to the sample being normal to the flow and, therefore, rotating. The regions of maximum area, instead, correspond to the sliding phase. According to these considerations, we assign the 0° angle to the first point minimizing the area of the specimen, as this point is always clearly identifiable, instead of assigning it to the first BM available. All the BMs before the first and successive to the last minimum value of the area correspond to an ambiguous assignation and are therefore discarded. By examining the other points of minimum area, the rotation halves can be identified, i.e., the projections corresponding to a flipping of the sample (180°) or a full rotation (360°).

In order, to identify the BMs corresponding to rotation quarters (90°, 270°, etc.), the ambiguity corresponding to the maximum area projections should be solved. The ensemble of BMs corresponding to the maximum area is identified by fixing a threshold over the area plot equal to the 90% of the maximum area value. Among all the candidates for the rotation quarter projections, only one BM is kept, corresponding to the middle point between two consecutive rotation halves, marked in black in Figure 5c. All the other candidate BMs are eliminated as they are redundant. The resulting exploited projections are indicated in red in Figure 5d, where the rotation quarters are now evident. It is worth underlying that this procedure is completely automatic, just like the one for spherical samples in [15]. This is remarkable in terms of repeatability and throughput of the tomographic reconstruction of flowing samples. To finally construct the angular sequence, the same method presented in Section 2.2 is used for each quarter of rotation. This method is particularly robust because it considers different cases. Indeed, in the case of symmetrical samples, the same angular speed is obtained for each quarter, while for asymmetrical samples (for example, replicating samples consisting of two asymmetric halves) the rotational speed detected for each quarter might be different. As a final consideration, the proposed angle tracking method does not rely on the phase values, as this would result in a method not applicable to samples for which, due to the severe scattering and or absorption, the phase estimation is uncertain.

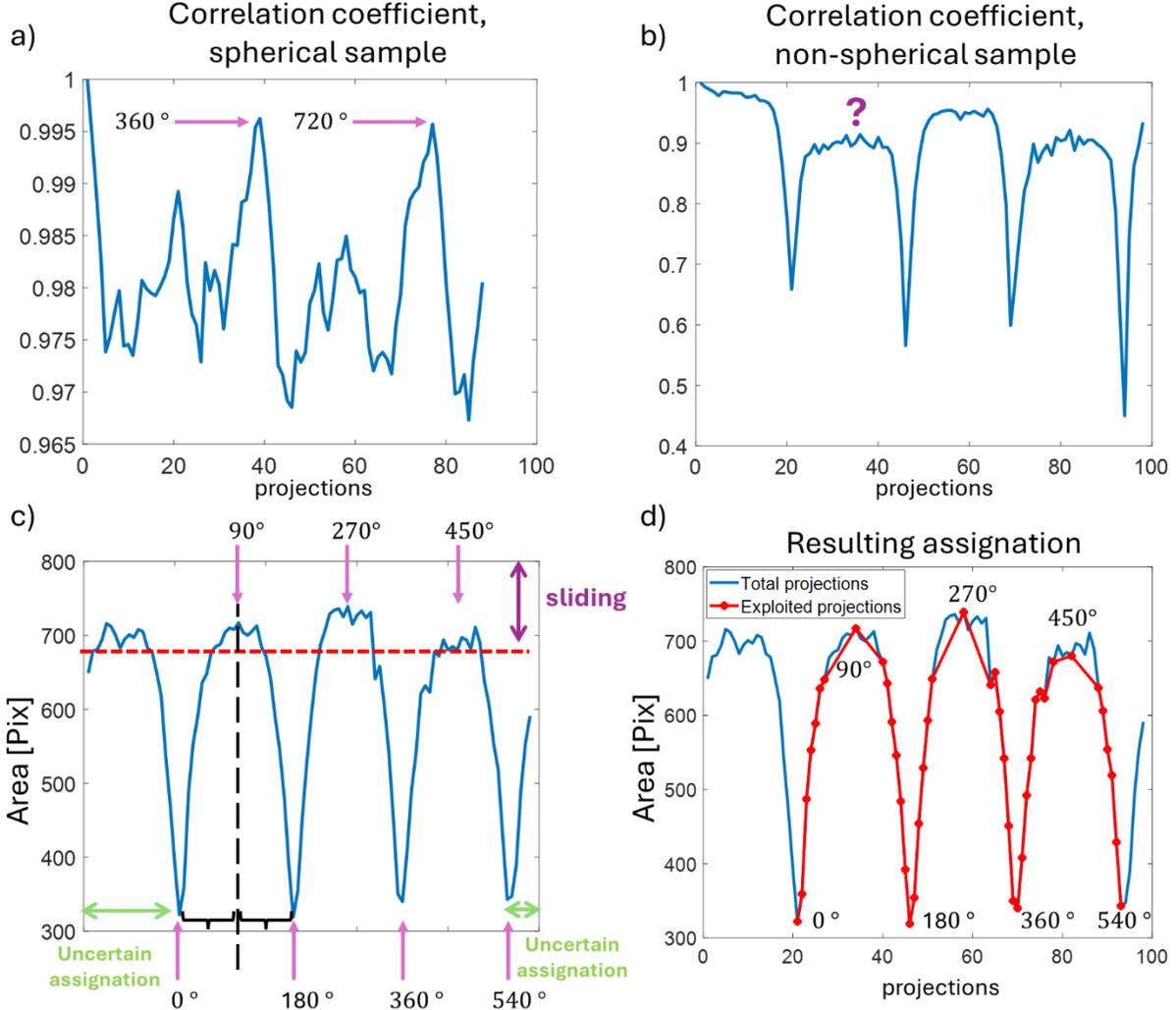

**Figure 5:** proposed angle retrieval method for non-spherical samples for Rotation pattern 1. (a), application of the method in [15] to a typical spherical object and (b) to a non-spherical sample (specifically a monocyte). (c) proposed method for the elimination of the redundancies in the sliding phase and the identification of the rotation quarters. (d) resulting BM sequence for the evaluation of the projection angles.

## 3 Results and discussions

Exploiting the angle retrieval approaches presented above, we reconstructed the 3D shape of several microalgae, specifically the diatoms *Cocconeis* sp., *Skeletonema pseudocostatum, Skeletonema marinoi*, *Navicula* sp. and the dinoflagellates *Scrippsiella acuminata*, *Heterocapsa* sp. and *Prorocentrum* sp., all belonging to the Rotation pattern 1 class, and the diatoms *Cocconeis* sp., *Thalassisira eccentrica and Skeletonema marinoi* belonging to the Rotation pattern 2 class. Details about the sample's preparation can be found in the Supplementary materials. For most diatoms we could perform tomography due to a sufficient quality of the acquired QPMs, therefore the entire 3D RI profile of the algae was reconstructed. For all the dinoflagellates and *Thalassisira eccentrica*, due to excessive absorption and excessive scattering by the frustule, respectively, we computed SFS. The angle retrieval approach presented, indeed, is suitable both for tomography and any other 3D reconstruction method like SFS. The differentiation between the two rotation patterns described in Section 2.1 was performed automatically evaluating the area of the sample in the sequence of QPMs. The samples for which a sharp decrease in area was followed by several frames with almost no area variation were assigned to Rotation pattern 2, while the samples for which a relatively small area variation frame to frame was calculated were assigned to Rotation pattern 1. Additional details regarding the effects of the application of the angle retrieval pipeline for spherical samples for Rotation pattern 2 are given in the Supplementary material, along with the effects of a missing compensation of the *apparent kayaking*.

*3.1 Tomography for Rotation Pattern 1 samples*

We retrieved the sequences of rolling angles of various microalgae exhibiting weak scattering, and thus we were able to reconstruct successfully the respective tomograms. Such diatoms experienced Rotation pattern 1 according to our nomenclature. Specifically, we report the 3D RI tomograms of a *Navicula sp.,* a chain of two *Skeletonema pseudocostatum* samples, a chain of two *Skeletonema marinoi* cells, a single cell of *Cocconeis* sp. and a cluster of two cells of *Cocconeis* sp. samples rotating jointly. To compute the tomograms, the High Order Total Variation (HOTV) of order = 2 was used as described in [23]. As a result, we present five examples of 3D RI tomograms of five microalgae species exhibiting the Rotation Pattern 1 (Figure 6). The central xz slice of the tomograms is displayed, together with the isolevel visualization obtained with three thresholds set as the 10%, 20% and 50% of the maximum value of RI, respectively. The peculiar internal structure of each diatom is retrieved (Figure 6). For *Navicula* sp., the two distinctively shaped chloroplasts are visible along the borders of the diatom, together with the low-RI central nucleus. Silicate structures such as the *raphe* cannot be visualized in the transmission RI measurement. The peculiar morphology of *Cocconeis* sp. is reconstructed, characterized by a thin elliptical shape with a high internal RI region associated to the chloroplast. The tomograms in Figure 6 have been calculated through the correct angle sequence estimation that compensate the compensation of the *apparent kayaking* and in fact the 3D tomograms have the expected morphology, while in case of missing the correction for kayaking the morphology appears to be clearly aberrated (see suppl. Figs S1 a.d). In the Supplementary material, we discuss the severe effects of the lack of compensation of the *apparent kayaking*. Supplementary Figure S6 summarizes the rotation behavior of some of the microalgae shown in this section.

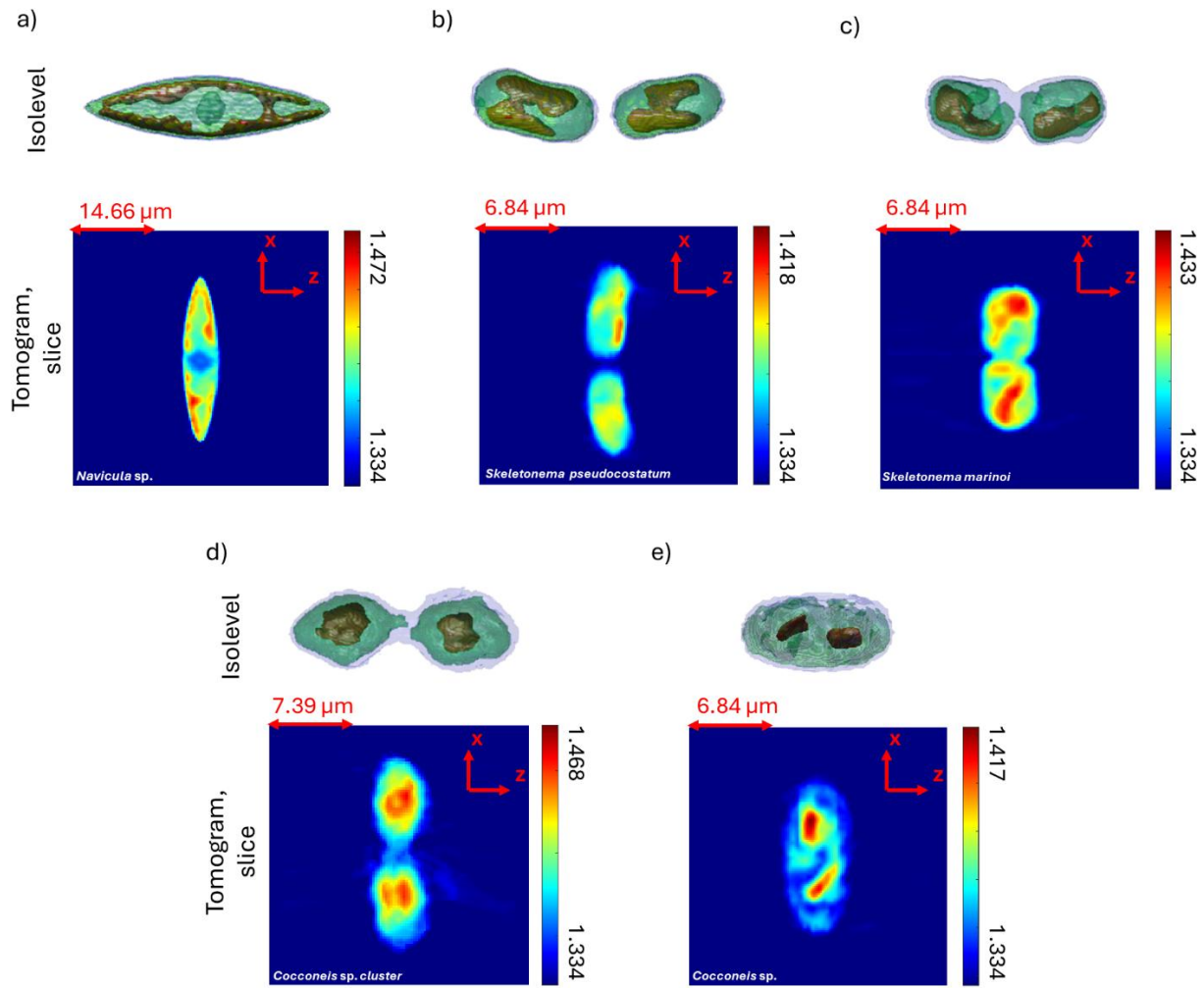

**Figure 6:** typical tomographic reconstruction of diatoms following Rotation pattern 1. Up, isolevel visualization, down, central slice of the tomogram. (a), *Navicula* sp. (b), *Skeletonema pseudocostatum*, (c), *Skeletonema marinoi*. (d), *Cocconeis* sp. cluster, (e), *Cocconeis* sp.

*3.2 Tomography for Rotation Pattern 2 samples*

Figure 7 illustrates the tomograms computed for diatoms belonging to the Rotation pattern 2 class. In this case, the diatoms reconstructed are again three cells of *Cocconeis* sp. and one chain of *Skeletonema marinoi*. However, due to a different orientation with which they entered the microfluidic channel, the rotation pattern exhibited is different from the case described above. This means that, unless a precise engineering of the fluid in the microfluidic channel is made, the orientation of the samples entering the channel is crucial to determine their rotation behavior, and therefore it is not possible to use the same angle retrieval approach for each sample without a previous assessment. The detrimental effects of a non-correctly retrieved angular sequence are highlighted in the Supplementary material in Figure S1 e-h. Due to their thin disk-shaped structure, the cells of *Cocconeis* sp. typically rotate with Rotation pattern 2, exhibiting the characteristic sliding – thumbling behavior. Their motion is often turbulent and non-stable among the consecutive 360° rotations in the FOV. The chains of *Skeletonema marinoi* tend to show the tumbling-sliding motion pattern too in most of the cases, as we assessed experimentally, often performing one single 180° flip in the available FOV. However, differently from *Cocconeis* sp., they tend to tumble in a less abrupt way, while remaining in the sliding phase for a bigger number of frames. Due to the above considerations, regarding the possibility to correct the spurious rotation components, in this case, only tomograms of samples not presenting the apparent kayaking rotation were computed. Generally, the approximation of the tomograms obtained is worse compared to Rotation pattern 1 case. However, using the proposed correct angle retrieval approach, it is still possible to compute tomograms that retrieve the actual sample's shape, and where the internal structures and the mapped expected RI range are in agreement with expectations. We have experimentally isolated a dead *Cocconeis* sp. cell and reconstruct its tomogram, in which the cytoplasm extrusion was visible (Figure 7a). Only the external membrane was left, and the chloroplasts collapsed together forming a round structure. These results further highlight the capability of the method to discern in diatoms signs of stress occurring, e.g., as a consequence of non-ideal environmental conditions. Supplementary Figure S6 summarizes the rotation behavior of some of the samples shown in this section.

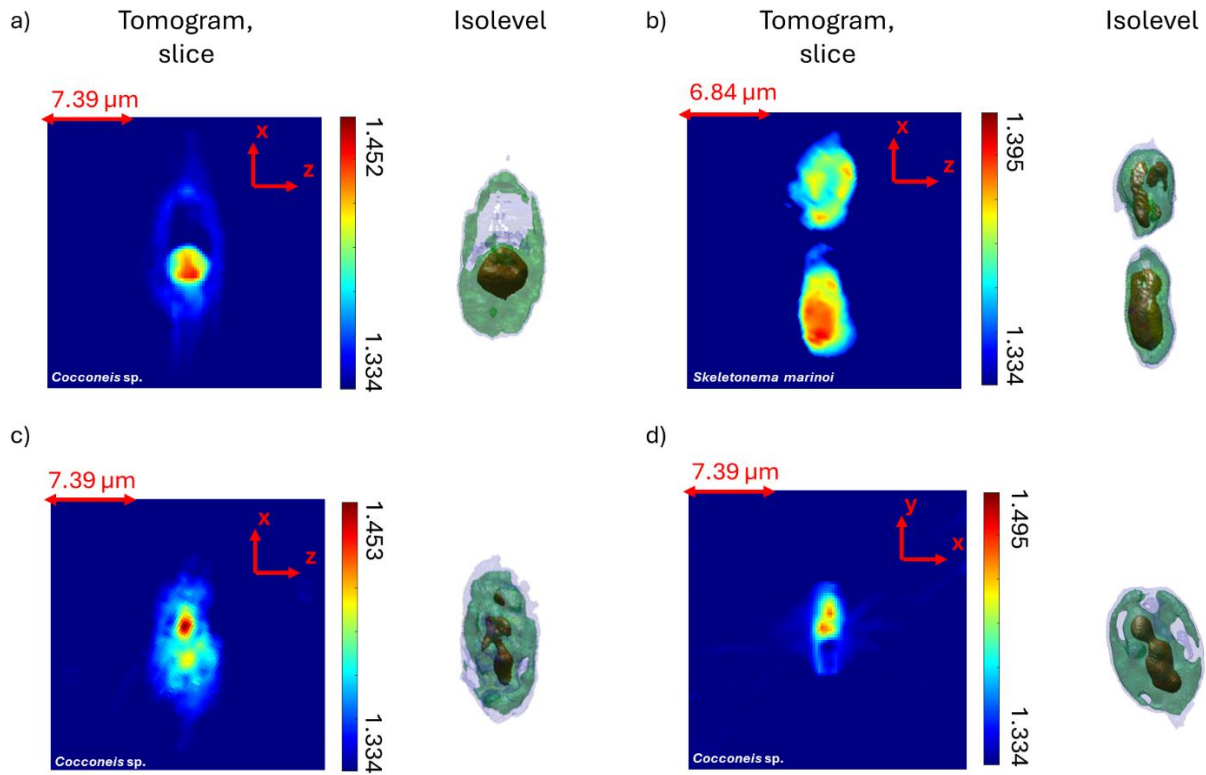

**Figure 7:** Tomographic reconstruction of diatoms following Rotation pattern 2. Right, isolevel visualization, left, central slice of the tomogram. (a), (c), (d), *Cocconeis* sp. (b), *Skeletonema marinoi*.

3.3 Shape From Silhouette for Rotation pattern 1 and 2 samples

We performed the SFS reconstruction of four additional species exhibiting Rotation pattern 1 or 2, and that were not suitable for FHT due to their optical properties. Specifically, we reconstructed *Scrippsiella acuminata*, *Heterocapsa* sp., *Prorocentrum* sp. and *Thalassiosira eccentrica*, shown in Figure 8a. From all the available 3D reconstructions of microalgae shown in this work, we extracted morphological parameters of interest (Table 1). The parameters were measured directly from the computed SFS (Figure 8a), or from the outer shells of the tomographic reconstructions, when available (Figure 8b). The intrinsic 3D nature of our reconstructions allows an access to the entire profile of the samples, accounting also for their anisotropy (Figure 8). Accordingly, the measurements of the parameters of interest and in particular of the biovolume do not rely on approximations of the 3D shape of the sample, differently from the state-of-the-art imaging tool for microalgae, namely the Imaging FlowCytobot (IFCB) [24] (See supplementary materials).

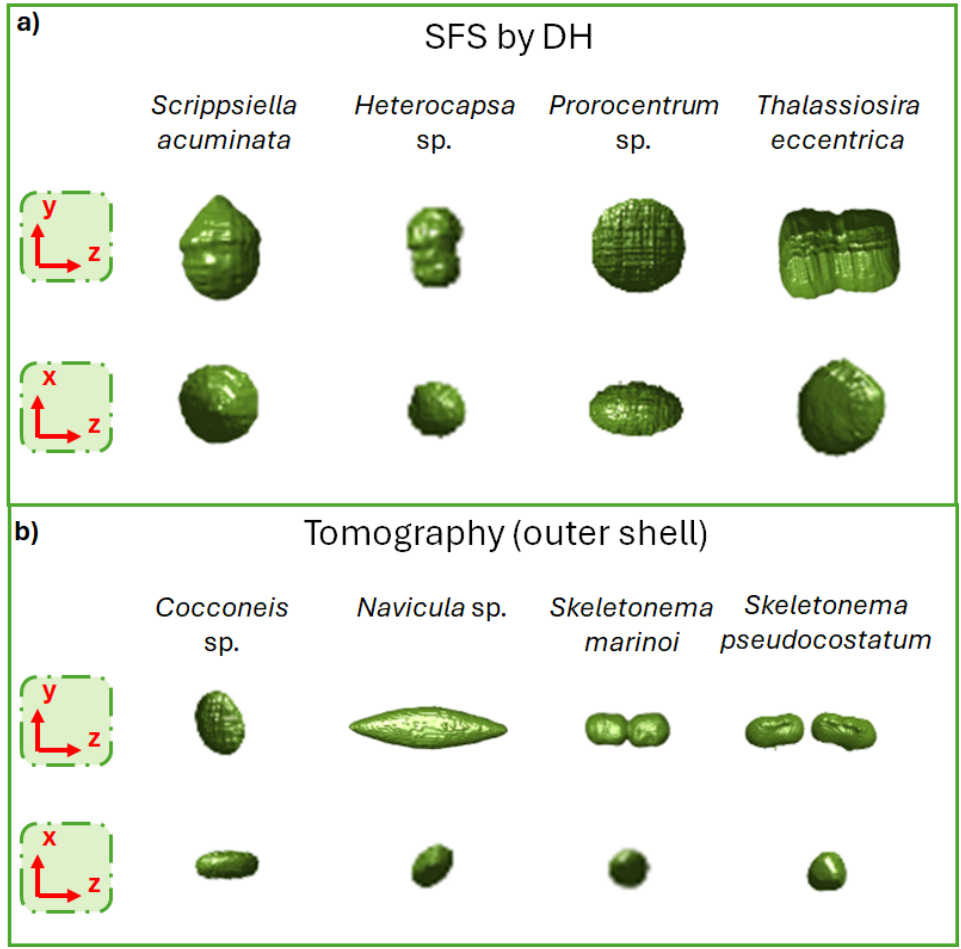

**Figure 8:** (a) SFS reconstructions of scattering or absorbing samples. (b), outer shells of tomographic reconstructions for a set of samples discussed in the previous sections.

| Sample | Volume [μm³] | Eccentricity | Equiv. Diameter [μm] |
| --- | --- | --- | --- |
| *Cocconeis* sp. | 97.61 | 0.663 | 6.71 |
| *Navicula* sp. | 443.91 | 0.97 | 13.41 |
| *Skeletonema marinoi* | 135.74 | 0.92 | 7.64 |
| *Skeletonema pseudocostatum* | 90.70 | 0.91 | 9.80 |
| *Scrippsiella acuminata* | 3079.02 | 0.615 | 18.42 |
| *Heterocapsa* sp. | 354.16 | 0.69 | 8.66 |
| *Prorocentrum* sp. | 10577.40 | 0.376 | 33.14 |
| *Thalassiosira eccentrica* | 14058.46 | 0.51 | 31.76 |

**Table 1:** parameters measured from the computed 3D reconstructions. When available, the parameters were measured over the outer shells of the tomographic reconstruction; otherwise, they were extracted from the SFS.

**Conclusions**

This work presented a workflow for the 3D reconstruction of non-spherical samples via in-flow holography, embedding a novel angle retrieval approach. The workflow considers the optical properties of the samples, reconstructing the 3D shape, starting from amplitude or phase-contrast measurements. The angle retrieval approach correctly considers the movement of the samples in the microfluidic environment and allows to effectively recover the projection sequence in different rotation conditions, thus allowing the 3D reconstruction that instead is missed by commercial systems working in 2D mode. As proof of concept, this method was employed for the 3D reconstruction of microalgae cells. Microalgae are marine microorganisms of prime interest due to their critical ecological roles, but they are also a challenging group for in-flow 3D imaging because of their wide range of morphologies, which induce complex rotational patterns. Although flowing cells of microalgae have been imaged in 3D before [14, 25], we are the first to show that 3D quantitative measurements can be obtained for a wide variety of microalgae presenting different shapes and exhibiting various uncontrolled rotation patterns (summarized in Supplementary Figure S6). In fact, we demonstrate that a single holographic system can retrieve 3D data independently of the shape, size, and intrinsic optical properties of the cells. The proposed workflow extends the variety of objects that can be measured in 3D. It could be used in future as an extensive tool to characterize in 3D the structure of microalgae by extracting relevant parameters such as the biovolume, allowing a step forward with respect to the state-of-the-art of the 2D imaging tools for marine microorganisms.

**Acknowledgments**

This work was supported by Progetto PRIN 2022 PNRR P2022MA95R- microFluidic platfOrm fOr The toMographic phAse micRoscopy of phytoplanKton assemblageS (FOOTMARKS)-CUP B53D23023980001. Finanziato dall'Unione Europea-NextGenerationEU.

Anna Italiano and Emanuele Somma aided the culture and SEM taxonomical identification of *Cocconeis* sp. diatoms, respectively. SEM imaging was provided by Franco Iamunno (SZN microscopy service @ RIMAR department).

L.R. acknowledges supporting by the National Biodiversity Future Centre (NBFC) Program, Italian Ministry of University and Research, PNRR, Missione 4 Componente 2 Investimento 1.4 (Project: CN00000033).

The authors thank Wiebe Kooistra and Urban Tillman for providing some of their cultures.

**References**

[1] E. B. Barbier, "Coastal ecosystem-based management," Ecological Applications 15(6), 1869–1873 (2005), https://doi.org/10.1890/04-0922.

[2] F. Azam, T. Fenchel, J. Field, J. Gray, L. Meyer-Reil, and F. Thingstad, "The ecological role of water-column microbes in the sea," Mar. Ecol. Prog. Ser. 10, 257–263 (1983), https://doi.org/10.3354/meps010257.

[3] European Union, "Directive 2000/60/EC of the European Parliament and of the Council of 23 October 2000 establishing a framework for Community action in the field of water policy," Off. J. Eur. Communities L 327, 1–72 (2000).

[4] S. D. Batten, W. W. Gregg, R. G. Henson, et al., "Global plankton monitoring," Global Ecology and Biogeography 28, 727–745 (2019).

[5] J. F. Tweddle, M. D. Tilstone, T. Smyth, et al., "Phytoplankton phenology in the Northeast Atlantic," Remote Sens. Environ. 215, 1–14 (2018); E. R. Eriksen, J. G. Paasche, L. Strand, et al., "Monitoring marine microplastics," Sci. Total Environ. 675, 1–12 (2019).

[6] L. Roselli, G. Caroppo, D. Stanca, et al., "Ecological status of transitional waters based on phytoplankton," Estuarine Coastal and Shelf Science 265, 107581 (2022). (EI Indexed)

[7] H. Utermöhl, "Zur Vervollkommnung der quantitativen Methodik," SIL Commun. 9(1), 1–38 (1958), https://doi.org/10.1080/05384680.1958.11904091.

[8] M. E. Sönmez, B. Altınsoy, B. Y. Öztürk, N. E. Gümüş, and N. Eczacıoğlu, "Deep learning-based classification of microalgae using light and scanning electron microscopy images," *Micron* 172, 103506 (2023). https://doi.org/10.1016/j.micron.2023.103506

[9] M. W. Fawley and K. P. Fawley, "Identification of eukaryotic microalgal strains," J. Appl. Phycol. 33, 971–984 (2021). https://doi.org/10.1007/s10811-020-02190-5

[10] I. Agnarsson and M. Kuntner, "Taxonomy in a changing world: Seeking solutions for a science in crisis," Syst. Biol. 56(3), 531–539 (2007), https://doi.org/10.1080/10635150701424546.

[11] A. Pierella Karlusich, et al., "Global distribution patterns of phytoplankton chlorophyll a: functional types and biogeography," Global Biogeochem. Cycles 36, e2021GB007235 (2022).

[12] V. Sonnet, L. Guidi, C. B. Mouw, G. Puggioni, and S. D. Ayata, "Length, width, shape regularity, and chain structure: time series analysis of phytoplankton morphology from imagery," Limnol. Oceanogr. 67, 1850–1864 (2022). https://doi.org/10.1002/lno.12147

[13] L. Miccio, G. Coppola, P. Memmolo, et al., "Holo-tomographic flow cytometry: a new paradigm in diagnostics by high-throughput and stain-free single-cell imaging," Proc. SPIE 12464, Optical Biopsy XXI, 124640I (2023), https://doi.org/10.1117/12.2675702.

[14] F. Merola, L. Miccio, P. Memmolo, et al., "Tomographic flow cytometry by digital holography," Light Sci. Appl. 6(4), e16241 (2017), https://doi.org/10.1038/lsa.2016.241.

[15] P. Memmolo, F. Merola, L. Miccio, et al., "3D morphometry of red blood cells by digital holography," Appl. Opt. 59(4), A65–A73 (2020), https://doi.org/10.1364/AO.404376.

[16] D. G. Lowe, "Distinctive image features from scale-invariant keypoints," Comput. Vis. Image Underst. 110(3), 346–359 (2008), https://doi.org/10.1016/j.cviu.2008.02.006.

[17] S. J. D. Prince, Computer Vision: Models, Learning, and Inference, Cambridge University Press (2012).

[18] R. A. Fisher, "On the mathematical foundations of theoretical statistics," Proc. R. Soc. Lond. A 222(1149), 309–368 (1922), https://doi.org/10.1098/rspa.1922.0078.

[19] G. Popescu, Y. Park, N. Lue, et al., "Optical imaging of cell mass and growth dynamics," Micromachines 12(3), 277 (2021), https://doi.org/10.3390/mi12030277.

[20] J. J. Ferraro, C. R. Moncayo, A. J. Berger, "Label-free optical imaging of single-cell morphology and dynamics," Rev. Sci. Instrum. 93, 083702 (2022), https://doi.org/10.1063/5.0100963.

[21] Y. C. Eldridge, A. J. Walsh, et al., "High-throughput single-cell analysis using microfluidics and optical imaging," Lab Chip 13, 4463–4473 (2013), https://doi.org/10.1039/C3LC50515D.

[22] Y. Kim, S. Shim, J. S. Park, et al., "Deep learning-based label-free cell classification using refractive index tomograms," Light Sci. Appl. 10, 153 (2021), https://doi.org/10.1038/s41377-021-00626-2.

[23] M. P. Matos, J. R. dos Santos, et al., "Quantification of circulating tumor cells using AI and label-free imaging," Cells 11(16), 2591 (2022), https://doi.org/10.3390/cells11162591.

[24] H. M. Sosik and R. J. Olson, "Automated taxonomic classification of phytoplankton using imaging-in-flow cytometry," Limnol. Oceanogr. Methods 10, 278–294 (2012), https://doi.org/10.4319/lom.2012.10.278.

[25] K. Umemura, Y. Matsukawa, Y. Ide, and S. Mayama, "Label-free imaging and analysis of subcellular parts of a living diatom Cylindrotheca sp. using optical diffraction tomography," MethodsX 7, 100889 (2020). https://doi.org/10.1016/j.mex.2020.100889

[26] H. Sosik, "ifcb-analysis," GitHub repository, 2022. [Online]. Available: https://github.com/hsosik/ifcb-analysis.

[27] E. A. Moberg and H. M. Sosik, "Distance maps to estimate cell volume from two-dimensional plankton images," Limnol. Oceanogr. Methods 10, 278–288 (2012). https://doi.org/10.4319/lom.2012.10.278

**Supplementary material**

**IFCB characterization of microalgae**

IFCB was used to characterize the populations discussed in this work. The IFCB is the state-of-the-art imaging tool commonly used for marine microorganisms. It captures at high throughput 2D in-focus bright-field images of flowing microalgae and extracts from them a collection of morphological parameters. Most of them are related to 2D measurements, such as area, values of the principal axes, or shape factors. However, the instrument can provide estimations of 3D parameters such as the biovolume [24], computed by generating a rotation solid starting from a single acquired image. Two milliliters of each culture were analyzed with the IFCB® configured to capture images by triggering on chlorophyll a fluorescence (PMT B = 0.65, Trig. B = 0.13). All image processing and feature extractions were performed by following the instructions for blob and feature extraction (V4) and using the 'IFCB analysis' MATLAB code package, both publicly available on GitHub [26]. These codes implement the data pipeline described in [24] and [27], which extracts size features from the images of each microalgae cell. Briefly, edge detection and boundary segmentation algorithms were applied on the original grayscale images to extract the boundary of the largest target of interest (the "blob"). From this blob, a set of size features – including major and minor axis length, area, filled area, perimeter, equivalent spherical diameter, eccentricity, solidity, ratio of area to squared perimeter - were measured using the MATLAB Image Processing Toolbox's *regionprops* function. Biovolumes were derived from the images by using a distance map, as described in [27]. Finally, all the size features were converted from pixels to microns using a conversion factor of 2.77 pixels per micron, which accounts for the magnification and camera resolution of the IFCB. Histograms of the most relevant features are displayed in Figures S1 and S2 together with an example of the bright-field images provided by the instrument. Table s1 summarizes the extracted parameters.

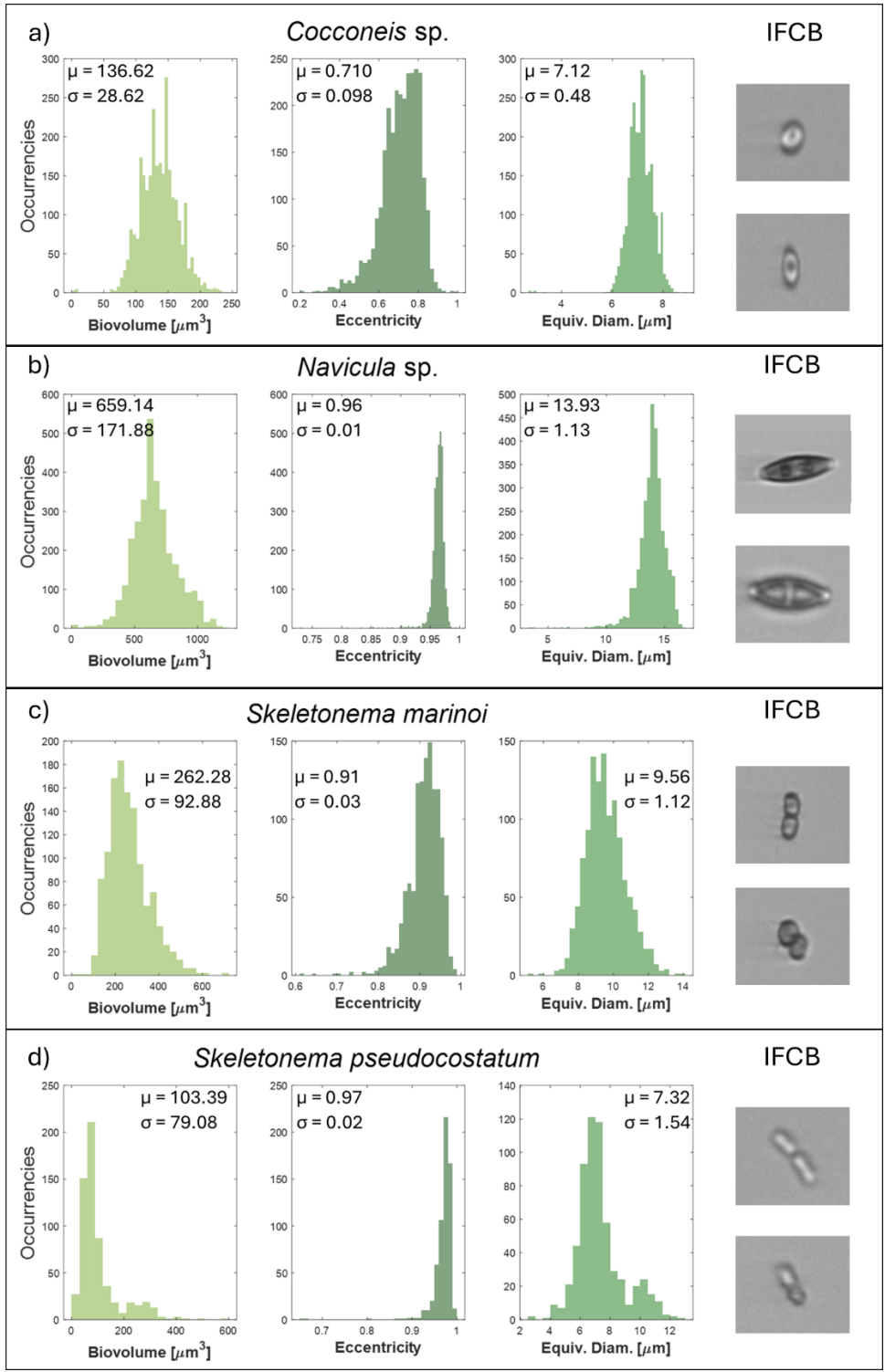

**Figure S1:** IFCB size features (biovolume, eccentricity and equivalent diameter) of (a), *Cocconeis* sp. (b), *Navicula* sp. (c), *Skeletonema marinoi* (d), *Skeletonema pseudocostatum.* μ = mean, σ = standard deviation.

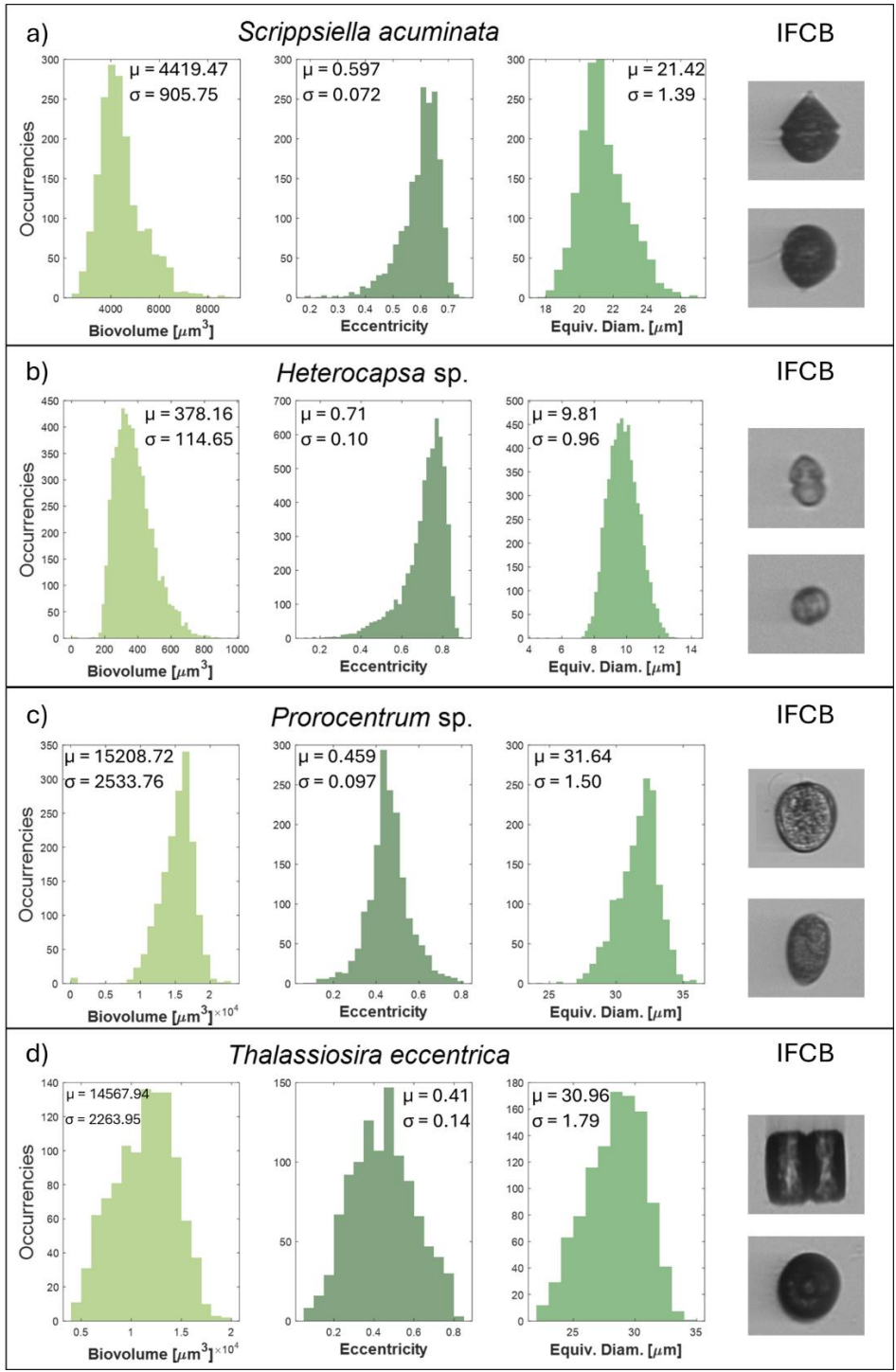

**Figure S2** IFCB size features (biovolume, eccentricity and equivalent diameter) of (a), *Scripsella acuminata*. (b), *Heterocapsa* sp. (c), *Prorocentrum* sp. (d) *Thalassiosira eccentrica*. µ = mean, σ = standard deviation.

|  | Area [$\mu m^2$] | Biovolume [$\mu m^3$] | Eccentricity | Eq. diam. [$\mu m$] | Extent [$\mu m^2$] | M. a. l [$\mu m$] | m. a. l [$\mu m$] |
|---|---|---|---|---|---|---|---|
| *Cocconeis* sp. | 40.02 ± 5.22 | 136.62 ± 28.62 | 0.71 ± 0.10 | 7.12 ± 0.48 | 0.09 ± 0.01 | 8.76 ± 0.83 | 6.00 ± 0.55 |
| *Navicula* sp. | 153.46 ± 23.41 | 659.14 ± 171.88 | 0.96 ± 0.01 | 13.93 ± 1.13 | 0.09 ± 0.01 | 27.61 ± 2.18 | 7.30 ± 0.74 |
| *Skeletonema marinoi* | 72.87 ± 17.36 | 262.28 ± 92.88 | 0.91 ± 0.03 | 9.56 ± 1.12 | 0.08 ± 0.01 | 15.93 ± 2.99 | 6.23 ± 0.78 |
| *Skeletonema pseudocostatum* | 43.92 ± 19.48 | 103.39 ± 79.08 | 0.97 ± 0.02 | 7.32 ± 1.54 | 0.06 ± 0.02 | 16.42 ± 2.99 | 3.78 ± 0.96 |
| *Scripsiella acuminata* | 361.83 ± 47.58 | 4419.47 ± 905.75 | 0.60 ± 0.07 | 21.43 ± 1.39 | 0.09 ± 0.01 | 24.14 ± 1.76 | 19.21 ± 1.36 |
| *Heterocapsa* sp. | 76.35 ± 14.90 | 378.16 ± 114.65 | 0.71 ± 0.10 | 9.81 ± 0.96 | 0.09 ± 0.01 | 12.01 ± 1.51 | 8.16 ± 0.86 |
| *Prorocentrum* sp. | 788.35 ± 73.56 | 15208.72 ± 2533.76 | 0.46 ± 0.10 | 31.65 ± 1.50 | 0.10 ± 0.00 | 33.81 ± 1.49 | 29.77 ± 1.90 |
| *Thalassiosira eccentrica* | 755.26 ± 90.34 | 14567.94 ± 2263.95 | 0.41 ± 0.14 | 30.96 ± 1.79 | 0.09 ± 0.01 | 33.20 ± 3.46 | 29.55 ± 1.31 |

**Table S1:** Size features measured with the IFCB (mean ± standard deviation). M. a. l=maximum axis length. m.a. l=minimum axis length

**Angle tracking for spherical particles rotating uniformly**

We briefly recall the procedure used for transparent spherical samples rotating uniformly as described in [15]. In our experimental setup the samples are injected in a rectangular microfluidic channel in which a laminar flow is established. For the samples flowing close to the channel bottom, rotation is induced since a torque is generated due to the non-even flow velocity applied at the margins of the cell. Being the samples well approximated by a sphere, the rotation obtained is homogeneous. As the samples flow and rotate into the microfluidic channel, digital holograms are recorded and numerically processed to extract the QPMs relative to different orientations. The starting points for the angular sequence estimation are the QPMs and the coordinates of the sample in the Field of View (FOV) for each orientation, obtained via a numerical tracking. Considering the stack of QPMs of a rotating sample, the first orientation is assigned without loss of generality to the angle 0°. Then, the projection corresponding to a complete rotation is identified by an opportune similarity metric, which allows to assign the 360° angle. The similarity metrics used leverage also the internal values of the QPMs, which are available due to the low scattering/absorption of transparent samples. Following, the entire angular sequence is estimated assuming a proportionality between the translation and the rotation, exploiting the coordinates of the sample in the FOV.

**Quality enhancement of tomograms**

Figure S3 illustrates the effects of a correct angle retrieval pipeline in terms of improvement of the tomographic reconstruction. Specifically, we investigate the consequences of the correction of the *apparent kayaking angle* for Rotation pattern 1 and the effects of supposing a proportionality between rotation and translation for Rotation pattern 2, neglecting the tumbling-sliding alternance. In Figure 10a, the *apparent kayaking* angle λ is identified on the QPMs of *Navicula* sp.; the corresponding λ-compensated QPMs are depicted in subfigure b. Figure 10 c and d show the central and a peripheral xy and xz slices of the tomograms obtained without and with the compensation of λ, respectively. As can be seen, if the compensation is not performed, the tomogram appears tilted and with artifacts characteristic of misalignment. Figure 10 e-h shows the effects of a wrong angle retrieval in the case of a *Cocconeis* sp. cell in Rotation pattern 2. In subfigure e, g two perspectives (lateral and frontal) of the isolevel and the central yz and xz slices of the tomogram, obtained using the proposed angle retrieval approach, are displayed. The corresponding images displayed in subfigures f and h were obtained with the angle retrieval approach for spherical samples introduced in [15]. The same thresholds were used for the isolevel. As can be seen, if a wrong angular sequence is reconstructed, the tomogram appears elongated, distorted, with incorrectly estimated RI values and shape. The thin ellipsoidal shape of *Cocconeis* sp. is not correctly reproduced, and the shape of the diatom appears rounder.

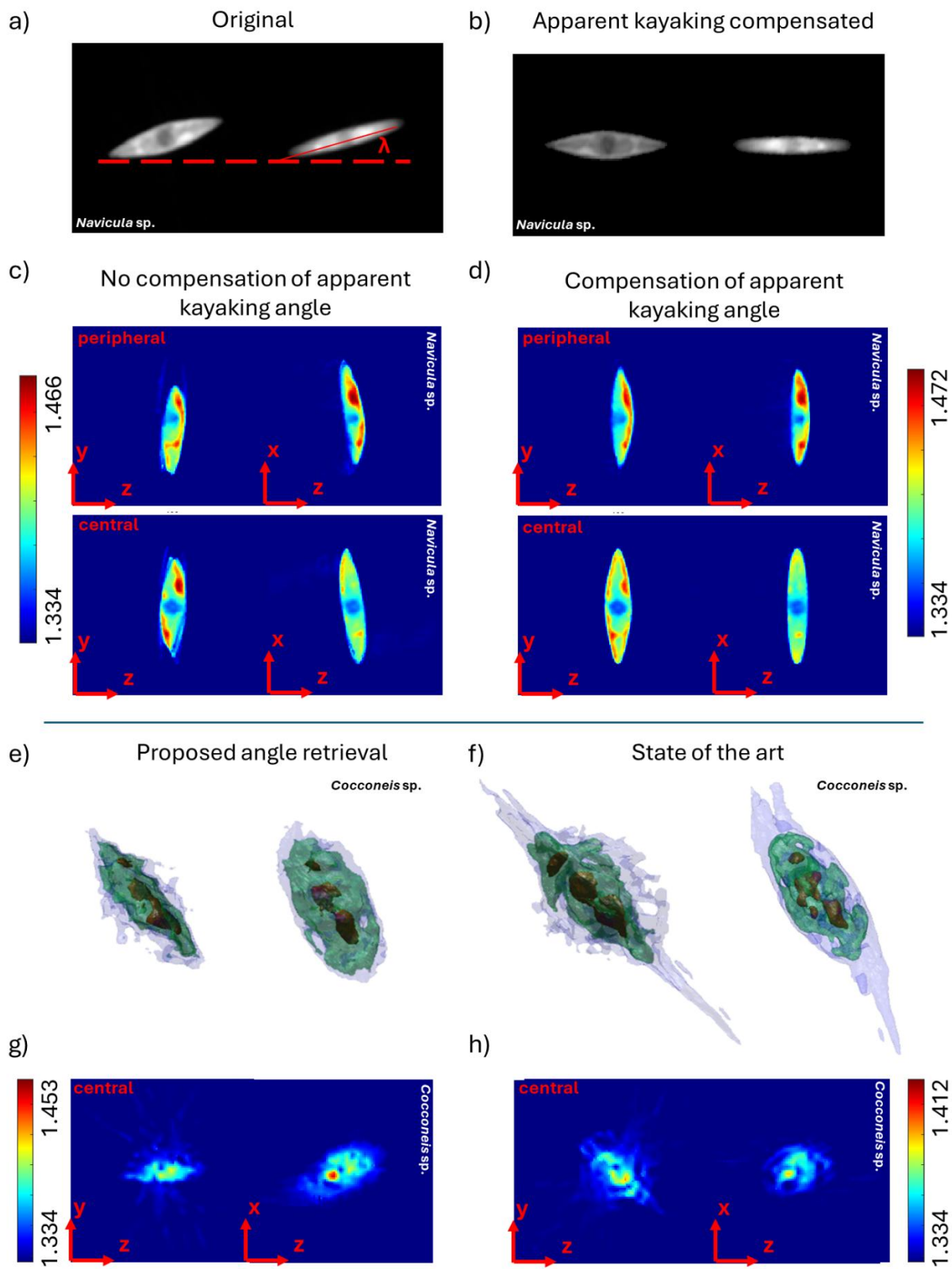

**Figure S3**: Effects of a correct angle retrieval in terms of quality improvement of the tomograms. Upper subpanel, effects of the apparent kayaking angle compensation. Lower panel, effects of the application of the angle retrieval

pipeline in [15] to a non-spherical sample. (a), apparent kayaking angle over the QPMs of a *Navicula* sp. (b), apparent kayaking angle compensated QPMs in (a). (c), peripheral and central slices of the tomogram of a *Navicula* sp. in case of no apparent kayaking angle compensation. (d), peripheral and central slices of the tomogram of a *Navicula* sp. in case of apparent kayaking angle compensation. (e), lateral and front profile of an isolevel visualization of the tomogram of a *Cocconeis* sp. obtained using the proposed angle retrieval method. (f), corresponding visualization of the isolevel obtained with the same thresholds in case of application of angle retrieval method in [15]. (g), central slice of the tomogram of the *Cocconeis* sp. in figure (e). (h), central slice of the tomogram of the *Cocconeis* sp. in figure (h).

**Isolation and culture conditions of microalgae**

Cultures of *Thalassiosira eccentrica*, and *Navicula* sp. were established from isolates collected in the Gulf of Naples (Tyrrhenian Sea) in 2024. The strain K2-C4 of *Scrippsiella acuminata* was isolated from the Danish Limford in 2020. All the cultures were maintained at 22°C under natural ambient light conditions, provided by window light. The culture medium was adapted to the species requirements. The diatoms *Thalassiosira eccentrica* and *Navicula* sp. were maintained in natural filtered seawater (salinity 36) autoclaved and supplemented with F/2 + Si. The dinoflagellates *Prorocentrum* sp., *Heterocapsa* sp. and *Scrippsiella acuminata* were maintained in natural filtered seawater (salinity 33) autoclaved and supplemented with K/2. Regarding the diatom *Skeletonema*, we used a culture of *S. pseudocostatum* (BS4), recently isolated from a coastal area facing the Sarno River mouth (40.72875N, 14.466432E, Naples, Italy), and two strains of *S. marinoi*, referred to as CCMP3318 and CCMP2092, collected from the Northern Adriatic Sea (sampling sites coordinates available at ncma.bigelow.org; access date: 31 March 2025). The *Cocconeis neothumensis* strains were isolated from two sampling sites off the Ischia Island (Naples, Italy). The first strain was isolated from the Cartaromana bay, near Sant'Anna rocks (40.433468N, 13.574092E), characterized by normal pH conditions (8.1), while the second strain was isolated off the Castello Aragonese (40.435062N, 13.574799E), a low-pH marine environment (7.6), due to the presence of $CO_2$ volcanic emissions. This pH, resembles that forecasted for our oceans in the year 2100. The *Skeletonema* strains (initial cell density: 5000 c/mL; working volume: 100 mL) were maintained in 75-cm$^2$ sterile flasks, filled with 100 mL of f/2 medium [1], under an irradiance of 100 µmol photons m$^{-2}$ s$^{-1}$ provided by 5000 k LED lamps, a 12: 12 light: dark photoperiod, and a temperature of 18°C. *Cocconeis* spp. were grown in glass Petri dishes (surface area: 23.75 cm²) using similar culture conditions with the exception of a higher irradiance (140 µmol photons m$^{-2}$ s$^{-1}$), provided by Sylvania GroLux neon lamps (Osram Sylvania Inc., USA). After 7 (*Skeletonema* spp.) days, 10-mL aliquots of each culture of planktonic algae were transferred in 15-mL falcon tubes and fixed with a Lugol iodine solution. In the case of benthic diatoms, after 15 days of culture cells were detached from the glass bottom using a steel blade, re-suspended in fixed volumes of sterile seawater, transferred in in 15-mL falcon tubes and fixed with a Lugol iodine solution. Cell density was estimated by counts in a Bürker chamber (Mannheim, Germany). The concentration of *Skeletonema* strains was around 600000-1600000 cells/mL, while the concentration of *Cocconeis* varied from 80000 (CnA) to 100000 (CnN) cells/mL. Algal species were diluted (adding appropriate amounts of culture medium) or concentrated in order to standardize the final cell density of all the samples, up to 400000 cells/mL. Then, 150 µl of each sample was transferred in 1.5-mL Eppendorf tubes and injected into the flow cytometry system.

**Optical setup**

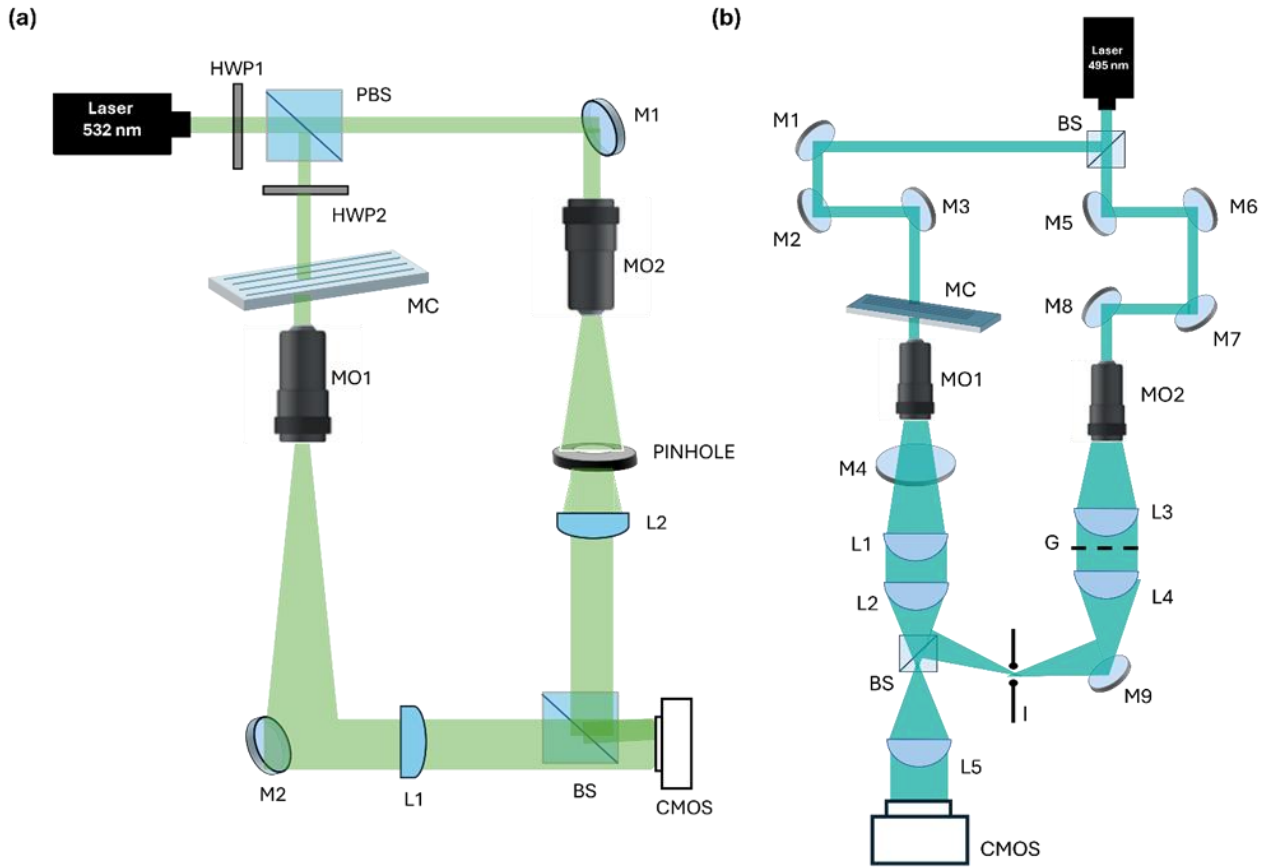

**Figure S4:** Schematic view of the FHT setups, where the light sources are a SuperK FIANIUM, 495 nm in (a) and Laser Quantum Torus, 532 nm in (b). BS, beam splitters; PBS, polarizing beam splitter; M, mirrors; MO, microscope objectives; L, lenses; G, diffraction grating; I, iris diaphragm; MC, microfluidic channel. HWP: Half-Wave Plate. CMOS: Complementary Metal-Oxide-Semiconductor.

To achieve holographic video recording, we implemented two distinct setups, both based on the Mach-Zehnder interferometer architecture. In the setup sketched in Figure S4a a continuous wave laser ($\lambda$=532 nm) generates a light, which is split by a polarizing beam splitter (PBS) into two different contributions (i.e., object and reference beams). The object beam passes through the flowing sample within the microfluidic channel (MC) and it is collected by an oil immersion microscope objective (MO1, 40x) with a numerical aperture equal to 0.9. Considering the wavelength of the used laser ($\lambda$=532 nm) and parameters of the microscope objective (NA=0.9), the maximum spatial frequency transmitted through the coherent imaging system is 1691 lp/mm which corresponds to a resolution of about 0.6 µm. At the same time, the reference beam passes through a beam expander shaped by a microscope objective and a second lens (MO2 and L2). A second beam splitter (BS) recombines the two contributions and the resulting interference fringe pattern is recorded using a CMOS camera (Genie Nano-CXP Cameras – 5120 × 5120 pixels, 4.5 µm pixel) at 30 fp. The overall lateral magnification in the camera plane reaches 36. The setup sketched in Figure S4b uses wave light at 495 nm. Microscope arrangements contain a microscope objective (MO2, 63x/0.95) without oil immersion in the signal arm, improving the maximum spatial frequency up to 1919 lp/mm, which corresponds to a resolution of about 0.5 µm. This setup enhances the total magnification to 76.4. To introduce a spatial carrier frequency essential for accurate off-axis phase retrieval, a diffraction grating (G) was strategically placed between the two lenses responsible for the reimaging process in the reference arm (l3 and L4 in Figure S4b). This approach also ensures that the spatial coherence between the interfering waves is maintained. Additionally, extra mirrors were integrated into the optical paths of both the reference and object beams, M1-M2 and M6-M7, in Figure S4b, respectively. These mirrors function as delay lines, effectively preserving temporal coherence, which is crucial given the finite spectral bandwidth of the employed light source. Diatoms *Cocconeis* sp., *Skeletonema marinoi*, *Skeletonema pseudocostatum*, *Navicula* sp. and *Thalassiosiora eccentrica* were acquired with the setup sketched in Figure S4b, while dinoflagellates *Heterocapsa* sp., *Scrippsiella acuminata* and *Prorocentrum* sp. were acquired with the setup sketched in Figure S4a.

**AM / QPM retrieval**

In the presented holo-tomographic tomographic flow cytometry system, a sample rotation paradigm is exploited. The samples are injected into a microfluidic chip where their rotation is induced by exploiting the parabolic velocity profile of the laminar flow generated in the channel. As they progress with their motion, digital holograms corresponding to different orientations of the specimens are recorded and stored. The digital holograms acquired are processed through a numerical pipeline devoted to reconstructing optically small, phase samples [23]. The principal steps of this pipeline comprehend 1) hologram segmentation, 2) hologram demodulation, 3) numerical refocusing, 4) amplitude/phase extraction, and, in case of availability of the phase, 5) aberration compensation, 6) phase unwrapping and 5) phase denoising. Step 1) allows to track and follow the sample into the Field of View (FOV) by numerically following its centroid as it flows into the FOV; In step 2) the diffraction order of interest is selected and filtered in the Fourier spectrum to obtain the out-of-focus complex amplitude. Step 3) is the numerical refocusing, an operation performed in post processing in which the hologram is numerically propagated up to the in-focus plane. In this work, the angular spectrum approach is chosen for the numerical propagation. The correct focus plane is individuated referring to a suitable contrast metric, which in our case is the Tamura coefficient. After this, in step 4) the in-focus amplitude or phase information is extracted. At the end of this step, the AMs are available. Step 5) allows eliminating the phase aberrations introduced by the optical system by exploiting a reference hologram, which is a sample-free hologram that only carries the aberration information. The reference hologram is subtracted in phase from the hologram to process in order to obtain an aberration-free image. As the obtained phase image is modulo $2\pi$ wrapped, a suitable unwrapping routine is exploited to retrieve the QPMs of the sample. Finally, in the last step, a denoising routine is performed over the QPMs to attenuate the speckle noise arising from the coherent light source used. In our case, the denoising routine is based on the windowed Fourier transform. All the parameters used in the described pipeline were fine tuned for the reconstruction of microalgae. Figure S5 shows a collection of QPMs relative to different diatoms, specifically the major and minor cross sections of *Navicula* sp. and *Cocconeis* sp., illustrating the non-isotropic morphology, and *Skeletonema marinoi* and *S. pseudocostatum*.

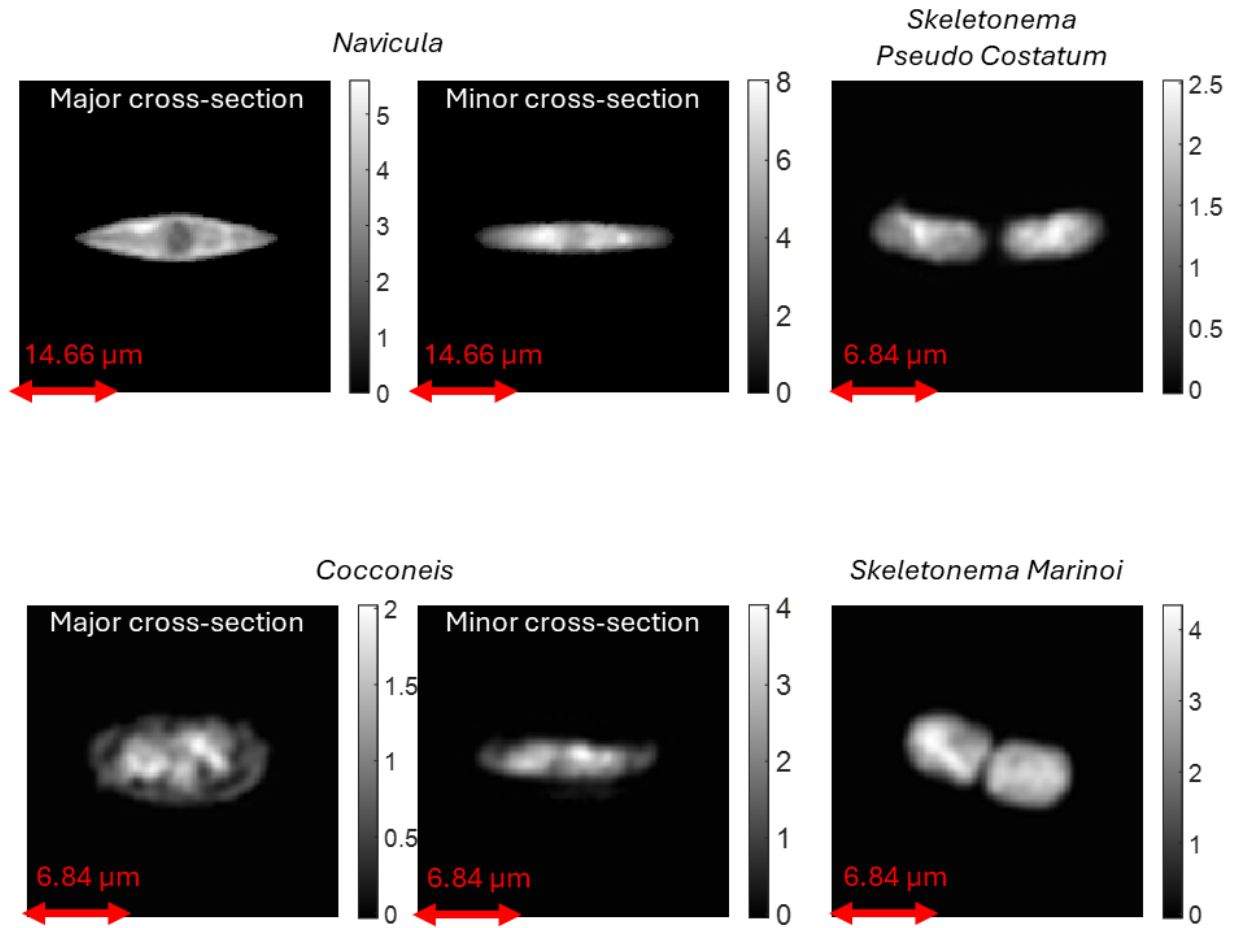

**Figure S5:** collection of obtained QPMs for different diatoms, specifically the major and minor cross sections of *Navicula* sp. and *Cocconeis* sp., illustrating the non-isotropic morphology, and *Skeletonema marinoi* and *pseudocostatum*.

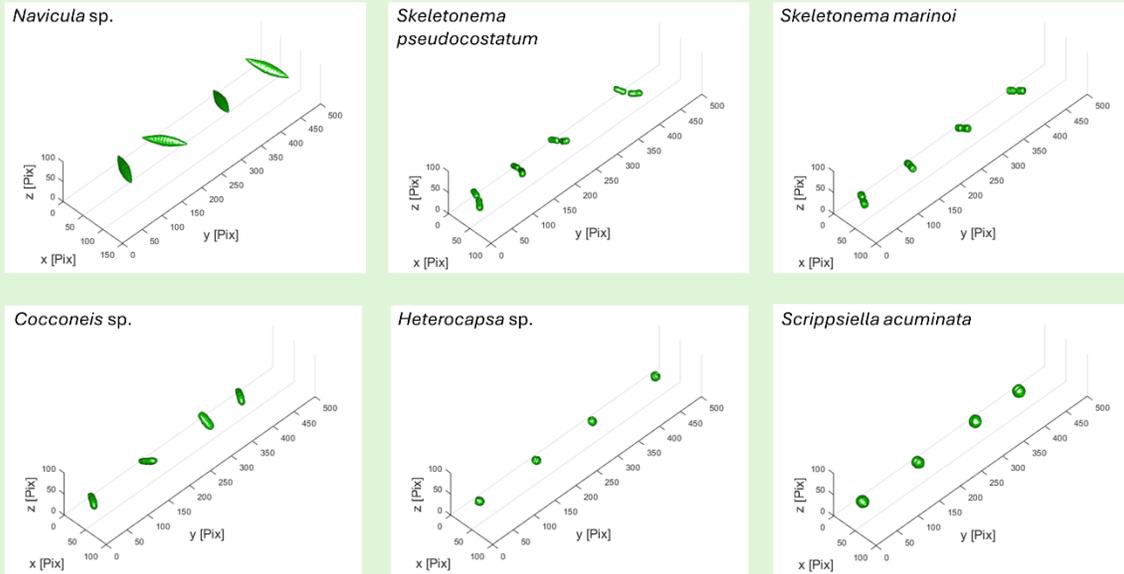
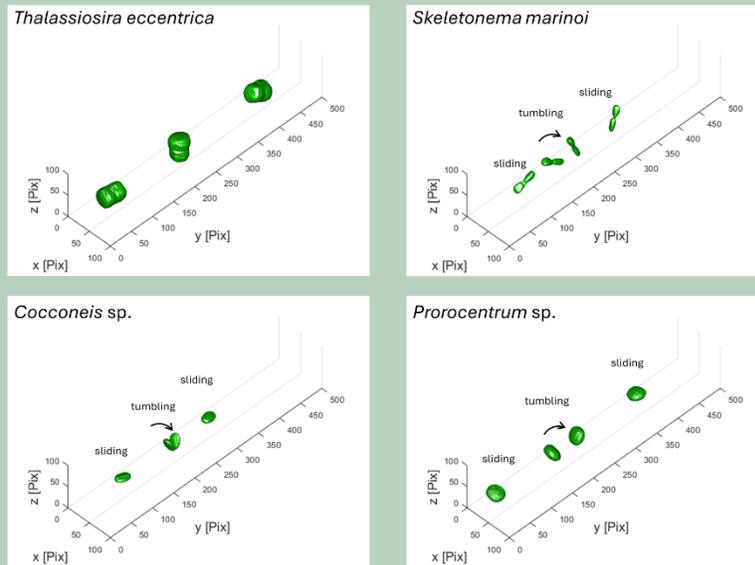

**Figure S6:** visualization of the rotation of the samples discussed in this work for Rotation patterns 1 and 2, obtained after 3D tracking and angle tracking. Different frames are superimposed to recall the movement in the microfluidic channel. For Rotation pattern 2, the tumbling-sliding phases are highlighted, if present.